\documentstyle[preprint,prb,aps]{revtex}

\begin{document}
\draft

\title{Critical Exponents for the SC-Model in the Zero Sector}

\author{Rudolf A.\ R\"{o}mer and Bill Sutherland}

\address{Physics Department, University of Utah, Salt Lake City, UT 84112}

\date{November 30, 1993}
\maketitle

\begin{abstract}
In this paper, we continue our investigation of a one-dimensional,
two-component, quantum many-body system in which like particles interact
with a pair potential $s(s+1)/{\rm sinh}^{2}(r)$, while unlike particles
interact with a pair potential $-s(s+1)/{\rm cosh}^{2}(r)$.
For an equal number of particles of the two components, the ground state
for $s>0$ corresponds
to an antiferromagnet/insulator. Excitations consist of
a gapless pair-hole--pair continuum, a two-particle continuum with gap
and excitons with gap. For $-1<s<0$, the system has two gapless
excitations --- a particle-hole continuum and a two spin-wave
continuum.
Using finite-size scaling methods of conformal field theory, we calculate
the asymptotic expressions and critical exponents for
correlation functions of these gapless excitations at zero temperature.
The conformal structure is closely related to the Hubbard model with
repulsive on-site interaction.
\end{abstract}


\narrowtext

%
%

\section{Introduction}

We recently presented the exact solution to a one-dimensional, two-component,
quantum many-body system  of considerable complexity in the form of an
asymptotic Bethe Ansatz calculation.\cite{sr93}
The two kinds of particles are distinguished by a quantum number
$\sigma= \pm 1$, which may be thought of as either spin or charge.
The system is defined by the Hamiltonian
\begin{equation}
H = - \sum_{1\leq j \leq N} \frac{1}{2}
       \frac{\partial^{2}}{\partial x_{j}^{2}}
    + \sum_{1\leq j < k \leq N} v_{jk}(x_{j}-x_{k}),
\end{equation}
where the pair potential is
\begin{equation}
v_{jk}(x) = s(s+1)
 \left[
  \frac{1+\sigma_{j}\sigma_{k}}{2 {\rm sinh}^{2}(x)}
  -
  \frac{1-\sigma_{j}\sigma_{k}}{2 {\rm cosh}^{2}(x)}
 \right].
\end{equation}
We assume $s\geq -1$.
We call this the SC-model, for the sinh-cosh interaction.
Thus for $s>0$, like particles repel, while unlike
particles attract.  When like particles are near, the repulsive potential
increases as $1/r^{2}$, while for large separations, both potentials decay
exponentially with a decay length which we take as our length scale, and hence
unity. The potentials might usefully be thought of as a screened $1/r^2$
potential.
This system was first introduced by Calogero \cite{crm75}, who showed it to be
integrable. Sutherland \cite{s78} soon afterward showed that the system could
be exactly solved, and gave the solution for a single component system.
In the present paper, we continue our study of the SC-model with an
investigation of the correlation functions and their critical exponents,
at zero temperature, by methods of conformal field theory.

The concept of conformal invariance in one-dimensional (1D) quantum systems
at criticality constrains the possible asymptotic behavior of correlation
functions and allows a classification into universality classes,
distinguished by the value
of the central charge $c$ of the underlying Virasoro algebra.\cite{cft}
For models with short-range interactions and a gapless excitation
spectrum with a single Fermi velocity, we can determine both $c$ and the
critical exponents of correlators directly from finite-size corrections
to the ground state energy and the low-lying exited states.
In most cases, such models have been found to belong to the universality
class of the 1D Luttinger model,\cite{h8x} i.e.\ $c=1$, and the critical
exponents to vary as functions of the coupling constant of the corresponding
conformal theory.

Recently, various authors have extended these concepts to include
multi-component systems with different excitation velocities, such
as the Hubbard model.\cite{w89,fk90}
In general, one finds a $c=1$ Virasoro algebra for each critical degree
of freedom, i.e.\ each gapless excitation with a unique velocity.
It is then possible to construct the full theory as a semidirect
product of these independent algebras. Again, critical exponents may
be calculated from finite-size corrections but now they are functions of
a matrix of coupling constants.

In another recent development, the ideas of conformal field
theory have been applied to models with long-range interactions
such as the $1/r^{2}$ system.\cite{mz,ky91}
It turns out that one can no longer simply read off the value of the central
charge from the finite-size corrections to the ground state energy.
However, one may still calculate the correct critical exponents of the
asymptotics of the correlations functions from the finite-size scaling
behavior of the low-lying excitations.\cite{rs93}

Our paper is organized as follows:
In section \ref{sec-baszs} we briefly review the asymptotic Bethe Ansatz
solution of the SC-model in the zero sector as obtained in Ref.\
\onlinecite{sr93}.
Section \ref{sec-cacf} outlines the Luttinger liquid approach for long-ranged
models. We give arguments why the standard evaluation of conformal field
theory for $c$ fails for long-ranged models.
For $s>0$, there is only one gapless excitation corresponding to a single
$c=1$ conformal theory. For $-1<s<0$, however, there are two gapless
excitations with different velocities, so that we briefly review the main
formulas for a semidirect product of two $c=1$ Virasoro algebras.
In section \ref{sec-acfa} and section \ref{sec-acfr}, we give expressions
for the correlation functions and calculate their critical exponents
from the Bethe Ansatz equations for the $-1<s<0$ and the $s>0$ cases,
respectively. For simplicity, the two types of particles are assumed to be
either both bosons or both fermions, although a mixed bose/fermi system can
be studied along similar lines.
We close our paper with section \ref{sec-nitcs}, where we
briefly show how both the $-1<s<0$ and the $s>0$ cases fit together as
$s\rightarrow 0^{\mp}$.

%
%

\section{The Bethe-Ansatz Solution in the Zero Sector}
\label{sec-baszs}

Let us recall the results of Ref.\onlinecite{sr93}: We have
$N_{\downarrow}$ particles with $\sigma=-1$ and $N_{\uparrow}$ with
$\sigma=+1$, for a total of $N=N_{\downarrow}+N_{\uparrow}$ and
 $N_{\uparrow}\geq N_{\downarrow}$. The zero sector corresponds to
an equal number of up and down spins, i.e.\ $N_{\downarrow}=N_{\uparrow}$.
For $s>0$, pairs of
up-down spins bind into a variety of bound states, or pairs, which we
will label by $m=1, \ldots, M(s)$, where $M(s)$ is the smallest integer
larger than $s$. Let there be $N_{m}$ of each type. Unbound particles
correspond to spinons/ions in the spin/charge picture and there are
$N_{0}= N - 2 \sum_{1\leq m\leq M} N_{m}$ of these. Let us call
particles with spin down spin waves; there are $N_{-1}= N_{\downarrow}
- \sum_{1\leq m\leq M} N_{m}$ of these.

Imposing periodic
boundary conditions on the wave function and taking any particle,
pair or spin wave around a ring of large circumference, yields the
following set of Bethe-Ansatz equations.
\begin{displaymath}
L \eta_m {\bf k}_m = 2\pi {\bf I}_m({\bf k}_m)
 + \sum_{-1 \leq m' \leq M} \sum_{{\bf k}'_{m'}}
    \theta_{m,m'}({\bf k}_m - {\bf k}_{m'}),
\end{displaymath}
\begin{equation}
m= -1, 0, 1, \ldots, M.
\label{eqn-bae}
\end{equation}
Here the ${\bf I}_m({\bf k}_m)$ denote the set of quantum numbers
for each type of particle.
Depending on the parities of $N_{m}$ and the particle
statistics, the quantum numbers will be restricted to integer or
half-odd integer values.
Note, that for the spin waves, ${\bf I}_{-1}$ ranges only over
$1,\ldots,N_0$.
$\theta_{m,m'}({\bf k}_{m}-{\bf k}_{m'})$ is the phase shift for the
scattering of particles of
type $m$ with type $m'$ and has been calculated previously.\cite{sr93}
Note that $\theta_{mm'}(k) = -\theta_{mm'}(-k) = \theta_{m'm}(k)$.
Furthermore, we define
\begin{equation}
\eta_m = \left\{
\begin{array}{ll}
0,              & m=-1,\\
1,              & m=0,\\
2,              & m=1,2,\ldots,M(s)
\end{array}
\right.
\end{equation}
We can write the momentum and energy for a
solution of (\ref{eqn-bae}) as
\begin{eqnarray}
P &= &\sum_{-1 \leq m \leq M} \eta_m \sum_{{\bf k}_m}{\bf k}_m,\\
E &= &\frac{1}{2}
      \sum_{-1 \leq m \leq M} \eta_m \sum_{{\bf k}_m}{\bf k}_{m}^2
     -\sum_{1 \leq m \leq M} N_m \kappa_{m}^2.
\end{eqnarray}
Here $\kappa_{m}=s+1-m$ denotes the poles in the transmission and reflection
coefficients that give rise to bound states of type $m$.

For $0>s>-1$, there are no bound states and we will call this the
{\em unbound case} in the sequel.
We therefore have only two coupled equations for $N_{0}$ particles with
pseudo-momenta
${\bf k}_{0}=(k_{1}, \ldots, k_{N_{0}})$ and $N_{-1}$ spin waves with
rapidities ${\bf k}_{-1}=(\lambda_{1}, \ldots, \lambda_{N_{-1}})$.
\begin{equation}
\begin{array}{lllll}
L k_{j} &= &2\pi I_{j}(k_{j})
 &+ &\sum_{\alpha=1}^{N_{-1}}\theta_{0,-1}(k_{j}-\lambda_{\alpha})
 \nonumber \\ & &
 &+ &\sum_{l=1}^{N_0}\theta_{0,0}(k_{j} - k_{l}), \\
0 &= &2\pi J_{\alpha}(\lambda_{\alpha})
 &+ &\sum_{\beta=1}^{N_{-1}}\theta_{-1,-1}(\lambda_{\alpha}-\lambda_{\beta})
 \nonumber \\ & &
 &+ &\sum_{j=1}^{N_0}\theta_{-1,0}(\lambda_{\alpha}- k_{j}).
\end{array}
\end{equation}
The particle quantum numbers $I_{j}$ and the spin-wave quantum numbers
$J_{\alpha}$ are restricted by the parities of $N_{0}$, $N_{-1}$ and
the statistics of the particles to the following combination of integers
and half-odd integers:
If both spin-up and spin-down particles are bosons
\begin{eqnarray}
  I_{j}     &= &(N_{0}-1)/2\ ({\rm mod}\ 1), \nonumber\\
  J_{\alpha}&= &(N_{-1}-1)/2\ ({\rm mod}\ 1),
\label{eqn-srbcc}
\end{eqnarray}
whereas for fermions,
\begin{eqnarray}
  I_{j}     &= &N_{-1}/2\ ({\rm mod}\ 1), \nonumber \\
  J_{\alpha}&= &(N_{0}+N_{-1}-1)/2\ ({\rm mod}\ 1).
\label{eqn-srfcc}
\end{eqnarray}

In the thermodynamic limit, i.e.\ $L\rightarrow\infty$ with fixed
$d_{0}\equiv N_{0}/L$, $d_{-1}\equiv N_{-1}/L$, the ground state is a filled
Fermi sea characterized by the distribution function $\rho(k)$ of particles
and $\sigma(\lambda)$ of down-spins.
\begin{equation}
\begin{array}{llcll}
\rho(k) &= &\frac{1}{2\pi} &+
 &\frac{1}{2\pi} \int_{-C}^{C}
  \theta_{0,-1}'(k - \mu) \sigma(\mu) {\rm d}\mu \\
 & & &+
 &\frac{1}{2\pi} \int_{-B}^{B}
  \theta_{0,0}'(k - h) \rho(h) {\rm d}h, \\
\sigma(\lambda) &= & 0 &+
 &\frac{1}{2\pi} \int_{-C}^{C}
  \theta_{-1,-1}'(\lambda-\mu) \sigma(\mu) {\rm d}\mu \\
 & & &+
 &\frac{1}{2\pi} \int_{-B}^{B}
  \theta_{-1,0}'(\lambda-h) \rho(h) {\rm d}h.
\end{array}
\end{equation}
Here the prime denotes the first derivative.
The values of $B$ and $C$ are fixed by the following equations:
\begin{eqnarray}
\int_{-B}^{B} \rho(k) {\rm d}k &= &d_{0},\\
\int_{-C}^{C} \sigma(\lambda) {\rm d}\lambda &= &d_{-1} = d_{0}/2 - {\cal M},
\end{eqnarray}
where ${\cal M}= (N_{\uparrow}-N_{\downarrow})/2L$ is the magnetization
per unit length.
Let us now restrict our discussion to the zero sector, when
$N_{0}=N$, $N_{-1}=N/2$, ${\cal M}=0$ and the limit $C$ of the spin wave
distribution is $\infty$.
Then we can solve for the spin wave distribution by Fourier transform
in terms of the particle distribution, which we then substitute
into the particle equation, giving a single integral equation for the
distribution of particles $\rho(k)$:
\begin{equation}
\frac{1}{2\pi} = \rho(k) + \frac{1}{2\pi}
 \int_{-B}^{B} \theta'(k-h)\rho(h){\rm d}h.
\end{equation}
Here the kernel $\theta'(k)$ is given as
\begin{equation}
\theta'(k)= \theta'_{00}(k) -
 1/2 \int_{-\infty}^{\infty} {\rm d}t e^{i k t}
\frac{{\rm sinh}t(1+s)}{{\rm sinh}t {\rm cosh}ts}.
\label{eqn-kern}
\end{equation}
The excited states in the zero-sector are given by:
(i) Remove a particle from the ground state distribution, and place it
outside the limits; we call this creating a hole and a particle, and it
gives a two parameter continuum.
(ii) Remove a spin wave from the ground state distribution, and place it on
the line with imaginary part equal to $i$; we call this creating two spin
waves,
one with spin up and the other with spin down. It gives a two parameter
continuum of the type familiar from the Heisenberg-Ising model.\cite{yy}
Each of these two types of two-particle continua has a single Fermi
velocity.
Let us denote by $v_{0}$ the Fermi velocity of the first excitation and by
$v_{-1}$ the Fermi velocity of the second. As has been pointed out in
Ref.\ \onlinecite{sr93}, the two velocities are in general not identical.
The same is true of the Hubbard model with repulsive on-site interaction,
and we will later make extensive use of the conformal results obtained
for this model.\cite{w89,fk90}

For $s>0$, which we call the {\em bound case}, the ground state in the
zero sector consists of a spin fluid of type $m=1$, and thus spin $0$.
In the ground state, the $k$'s for the pairs distribute themselves densely
with a density $\tau(k)$, between limits $\pm D$, normalized so that
\begin{equation}
d_{1} \equiv N_{1}/L = \int_{-D}^{D} \tau(k) {\rm d}k = N/2L.
\end{equation}
The energy and momentum are given by
\begin{eqnarray}
P/L &= & 2 \int_{-D}^{D} \tau(k) k {\rm d}k = 0,\\
E/L &= &   \int_{-D}^{D} \tau(k) k^2 {\rm d}k - s^2 N_1/L.
\end{eqnarray}
The integral equation which determines $\tau(k)$ is
\begin{equation}
1/\pi = \tau(k) +
 \frac{1}{2\pi} \int_{-D}^{D} \theta'_{11}(k-h)\tau(h) {\rm d}h.
\end{equation}
The kernel of the equation, $\theta_{11}'(k)$, is the derivative of the
phase shift for pair-pair scattering.

The low-energy excited states are given by the following:
(i) Remove a pair from the ground state distribution, and place it outside the
limits; we call this creating a pair-hole and a pair, and it gives a
gapless two parameter continuum.
(ii)  Break a pair, to give two particles, one spin up and the other spin
down;  this also gives a two parameter continuum. However, there is a
finite energy gap for breaking a pair. These are the spinons or ions.
(iii)  Excite a pair into a higher energy bound state, if allowed; these we
call excitons, and they have simple single parameter dispersion relations.

Let us denote the unique velocity of the excitations of type (i) by $v_{1}$.
The Bethe-Ansatz equations that describe these excitations
may be written as
\begin{equation}
2 L k_{j} = 2\pi H_{j}(k_{j})
 + \sum_{l=1}^{N/2} \theta_{-1,-1}(k_{j} - k_{l}).
\end{equation}
Note here that $k_{j}$ is the pseudo-momentum of a pair, and is not the
pseudo-momentum of an individual particle, which would be complex and of
the form $k_{j}/2 \pm is$.
The pair quantum numbers $H_{j}$ are restricted by the parity of
$N_{1}$, and bose and fermi statistics are given as
\begin{equation}
  H_{j}= (N_{1}+1)/2\ ({\rm mod}\ 1),
\label{eqn-src}
\end{equation}
since pair-pair scattering is symmetric for pairs of bosons and
pairs of fermions. The pairs will be singlets.

%
%

\section{Conformal Approach for Correlation Functions}
\label{sec-cacf}

\subsection{finite-size scaling in conformal theories of Luttinger liquids}

The behavior of the correlation functions for a given one-dimensional model
at large distances and low-temperatures is determined by the gapless
excitations.\cite{el75} These gapless excitations are due to
hydrodynamic fluctuations and it has been argued \cite{h8x} that the low
energy physics of such a system may be described by the exactly solvable
Luttinger model,\cite{lm65} the 1D quantum version of the classical 2D
Gaussian model.
The Luttinger model is a critical system with continuously varying
exponents and corresponds to the universality class of $c=1$ conformal
field theories.\cite{c87}
Application of conformal theory allows the calculation of these critical
exponents purely from finite-size scaling arguments.\cite{c8x}

The value of the central charge $c$ may be
read off from the following finite-size scaling formula
\begin{equation}
  E_{0} \sim \epsilon_{0} L -\frac{\pi v}{6 L} c,
\label{eqn-c}
\end{equation}
thus enabling an independent check of the above arguments.
Here, $E_{0}$ is the ground state energy of the finite system,
$\epsilon_{0}$ is the ground state energy density in the thermodynamic
limit and $v$ is the Fermi velocity in the system.
In short-ranged 1D quantum models, including Bethe-Ansatz solvable models,
the above universal picture is confirmed.\cite{sr}
However, for long-ranged models, straightforward
application of this
equation may lead to unphysical results.\cite{ky91}
(we include the SC-model in this class, although its pair potential decays
exponentially, since it can only be solved by means of the {\em asymptotic}
Bethe Ansatz.) For instance, in the
$1/r^{2}$ models $c$ is predicted to be equal to the interaction strength,
although independent calculations show that the critical exponents are
those of the $c=1$ universality class.\cite{rs93}
However, if one instead estimates $c$ from the low temperature expansion
of the free energy,\cite{c} one does get the correct answer $c=1$.

We can, however, understand the failure of (\ref{eqn-c}) in long-ranged
models. The crucial point is that, due to the long-range character of the
interactions, finite systems will always ``feel'' the particular
boundary conditions chosen, so that (\ref{eqn-c}) includes an additional
correction term $E_{bc}$, representing the boundary energy, and so
\begin{equation}
  E_{0} \sim \epsilon_{0} L -\frac{\pi v}{6 L} c + \frac{E_{bc}}{L}.
\label{eqn-cbc}
\end{equation}
The low temperature expansion, however, uses boundary conditions instead
for the time axis of the model and we thus have no such corrections.
We may therefore write the free energy of a long-ranged $c=1$ Luttinger
liquid as
\begin{equation}
F(T) \simeq F(T=0) - \frac{\pi T^2}{6 v}.
\label{eqn-fe}
\end{equation}

Let us recall the main formulas for calculating the correlation
functions and their critical exponents.\cite{cft}
Every primary field $\phi_{\pm}$ in a
conformal field theory on an infinite strip of width $L$ in the space
direction gives rise to a tower of exited states. Let
$x=\Delta^{+} + \Delta^{-}$ denote the scaling dimension and
$\sigma= \Delta^{+} - \Delta^{-}$ the spin of $\phi_{\pm}$.
Then the energies and momenta of these exited states scale as
\begin{eqnarray}
E(\Delta^{\pm},N^{\pm}) - E_{0}
 &\sim & \frac{2\pi v}{L} ( x + N^{+} + N^{-} ),
 \label{eqn-sec}\\
P(\Delta^{\pm},N^{\pm}) - P_{0}
 &\sim & \frac{2\pi}{L}( \sigma + N^{+} - N^{-} ) + 2 D k_{f}.
 \label{eqn-spc}
\end{eqnarray}
Here $N^{+}$ and $N^{-}$ are positive integers, $v$ is the common Fermi
velocity of the excitations and $2 D$ is the momentum
of the state in units of the Fermi momentum $k_{f}$.
Note that the quantities on the left hand side of these equations are
evaluated with respect to the same boundary condition and therefore the above
mentioned correction terms cancel.
We may write the correlation functions of the primary fields at zero
temperature (expressions for low but finite temperature may also be given)
as
\begin{equation}
\langle
 \phi_{\Delta^{\pm}}(x,t)
 \phi_{\Delta^{\pm}}(0,0)
\rangle
=
\frac{ \exp(-2 i D k_{f}) }
{ (x-ivt)^{2\Delta^{+}} (x+ivt)^{2\Delta^{-}} }.
\end{equation}
However, the excitation spectrum of the SC-model is quite different for
the bound ($s>0$) and the unbound ($-1<s<0$) case as we have
argued in the previous section. Most importantly, the unbound case
does not have a common velocity for all excitations anymore
and so the formulas given above for a Lorentz-invariant conformal
field theory can no longer hold.

\subsection{conformal weights and the dressed charge in the bound case}

For the bound case in the zero sector, only the pair--pair-hole
excitation branch is gapless. Thus there is only one excitation velocity
and from the above arguments, we expect the dimensions of the
primary operators to obey the formulae for a single $c=1$ Gaussian model,
i.e.\
\begin{equation}
 \Delta^{\pm}( \Delta N_{1}, D_{1} )
 = \frac{1}{2} \left( D_{1} \xi_1 \pm \frac{\Delta N_{1}}{2\xi_1} \right)^{2}.
\label{eqn-sdc}
\end{equation}
The coupling constant $\xi_1$ of this Gaussian model depends on the
system parameters. It is sometimes called the dressed charge and
may be calculated from the Bethe-Ansatz equations by means of an
integral equation \cite{w89}
\begin{equation}
 \xi_1(k) = 2
  +\frac{1}{2\pi} \int_{-D}^{D} \xi_1(h) \theta'_{1,1}(h-k) {\rm d}h,
\end{equation}
where the constant is $2$ because this excitation is a pair.
However, we can also calculate $\xi_1\equiv\xi_1(D)$ by purely thermodynamical
arguments as follows:
Let us change a given ground state configuration by adding pairs
while keeping the Fermi sea at zero momentum, so that the excitation
can be described by the pair $(\Delta N_{1}, D_{1}=0)$. Then a
second order expansion gives
\begin{equation}
  \Delta E = -\mu_{1} (\Delta N_1)
    + \frac{1}{2} \frac{1}{L \kappa_{1}d_{1}^{2}} (\Delta N_{1})^{2},
\end{equation}
where $\mu_{1}= - \frac{\partial E}{\partial N_{1}}$ is the chemical potential
for adding pairs and $\kappa_{1}$ is the pair-compressibility.
Comparison with (\ref{eqn-sec}) and (\ref{eqn-sdc}) yields
\begin{equation}
  \xi_1^{2} = \pi v_{1} \kappa_{1} d_{1}^{2} = \pi d_{1}/v_{1}.
\label{eqn-xic}
\end{equation}
In the last equation, we have used the well known relation
$v_{1}^{2} = 1/(\kappa_{1} d_{1})$.
Therefore, by knowing the Fermi velocity of the pair -- pair-hole
excitations, we can calculate the scaling dimensions.

\subsection{finite-size scaling and the dressed charge matrix in the
unbound case}

For the unbound case, two excitation branches are gapless,
giving rise to a particle-hole continuum and to a spin wave continuum,
with Fermi velocities $v_{0}$ and $v_{-1}$, respectively.
Thus, the finite size corrections of equations (\ref{eqn-sec}) and
(\ref{eqn-spc}) now become
\widetext
\begin{equation}
E( \Delta{\bf N}, {\bf D} ) - E_{0}
 \sim \frac{2\pi}{L}
 \left[
  \frac{1}{4} \Delta{\bf N}^{T} (\Xi^{-1})^{T} V (\Xi^{-1}) \Delta{\bf N}
  + {\bf D}^{T} \Xi V \Xi^{T} {\bf D}
  + v_{0} ( N_{0}^{+} + N_{0}^{-} )
  + v_{-1}  ( N_{-1}^{+} + N_{-1}^{-} )
 \right],
\label{eqn-secc}
\end{equation}
\begin{equation}
P( \Delta{\bf N}, {\bf D} ) - P_{0}
 \sim \frac{2\pi}{L}
 \left[
  \Delta{\bf N}^{T}\ {\bf D}
  + N_{0}^{+} - N_{0}^{-}
  + N_{-1}^{+} - N_{-1}^{-}
 \right]
 + 2 D_{0} k_{f,\uparrow}
 + 2 ( D_{0} + D_{-1} ) k_{f,\downarrow}
\label{eqn-spcc}
\end{equation}
\narrowtext
Here, the matrix $V\equiv {\rm diag}(v_{0}, v_{-1})$ and the excited state is
characterized by the pairs $\Delta{\bf N}= (\Delta N_{0}, \Delta N_{-1})$
and ${\bf D}= (D_{0}, D_{-1})$.
As before, $N_{0}^{\pm}$ and $N_{-1}^{\pm}$ are positive integers
that label the descendant fields.
The $2\times 2$ matrix $\Xi$ is the generalization of the dressed
charge $\xi$ and may be calculated by means of coupled integral
equations. Thus if we denote the components of $\Xi$ by
\begin{equation}
\Xi = \left( \begin{array}{cc}
              \xi_{0,0}(B)    & \xi_{0,-1}(C) \\
              \xi_{-1,0}(B)   & \xi_{-1,-1}(C)
             \end{array}
      \right),
\end{equation}
then
\begin{equation}
\begin{array}{lclll}
 \xi_{0,0}(k) &= & 1
  &+&\frac{1}{2\pi} \int_{-B}^{B} \xi_{0,0}(h) \theta'_{0,0}(h-k) {\rm d}h
  \\ & &
  &+&\frac{1}{2\pi} \int_{-C}^{C} \xi_{0,-1}(\mu) \theta'_{-1,0}(\mu-k) {\rm
d}\mu.\\
 \xi_{0,-1}(\lambda) &= & 0
  &+&\frac{1}{2\pi} \int_{-B}^{B} \xi_{0,0}(h) \theta'_{0,-1}(h-\lambda) {\rm
d}h
  \\ & &
  &+&\frac{1}{2\pi} \int_{-C}^{C} \xi_{0,-1}(\mu) \theta'_{-1,-1}(\mu-\lambda)
{\rm d}\mu, \\
 \xi_{-1,0}(k) &= & 0
  &+&\frac{1}{2\pi} \int_{-B}^{B} \xi_{-1,0}(h) \theta'_{0,0}(h-k) {\rm d}h
  \\ & &
  &+&\frac{1}{2\pi} \int_{-C}^{C} \xi_{-1,-1}(\mu) \theta'_{-1,0}(\mu-k) {\rm
d}\mu, \\
 \xi_{-1,-1}(\lambda) &= & 1
  &+&\frac{1}{2\pi} \int_{-B}^{B} \xi_{-1,0}(h) \theta'_{0,-1}(h-\lambda) {\rm
d}h
  \\ & &
  &+&\frac{1}{2\pi} \int_{-C}^{C} \xi_{-1,-1}(\mu) \theta'_{-1,-1}(\mu-\lambda)
{\rm d}\mu.
\end{array}
\end{equation}
Thus, the situation for $-1<s<0$ is analogous to the situation in the
repulsive Hubbard model away from half-filling \cite{fk90,w89} and
we may interpret equations (\ref{eqn-secc}) and (\ref{eqn-spcc}) in
terms of a semidirect product of two independent Virasoro algebras,
both with $c=1$.
The scaling behavior of the energy and momentum in terms of the
conformal weights $\Delta_{0}^{\pm}$ and $\Delta_{-1}^{\pm}$ and the
formulas for these weights as functions of the components of the dressed
charge matrix $\Xi$ have been given in Ref. \onlinecite{fk90}, and
we will not repeat them here.
The generalization of the correlation functions of the primary fields
has also been given in Ref. \onlinecite{fk90}. However, as before,
thermodynamic arguments may be used to calculate the values of
the dressed charge matrix.

For the zero sector, i.e.\ ${\cal M}=0$, the relevant equations
simplify considerably. In this case,
$k_{f,\downarrow}=k_{f,\uparrow}\equiv k_{f}=\pi d_{0}/2$, and
the dressed charge matrix $\Xi$ may
again be expressed in terms of a single parameter $\xi_0\equiv\xi_0(B)$, i.e.\
\begin{equation}
\Xi = \left( \begin{array}{cc}
              \xi_{0}              & 0 \\
              \frac{1}{2}\xi_{0}   & \frac{1}{\sqrt{2(1+s)}}
             \end{array}
      \right).
\end{equation}
Thus the conformal weights $\Delta_{0}^{\pm}$ and $\Delta_{-1}^{\pm}$ are
given as
\begin{eqnarray}
\Delta_{0}^{\pm} &= &
 \frac{1}{2} \xi_0^{2}( D_{0} + \frac{1}{2} D_{-1} )^{2}
 + \frac{1}{8\xi_0^{2}} ( \Delta N_{0} )^{2}
 \nonumber \\ & & \mbox{ }\
 \pm \frac{1}{4} \Delta N_{0} ( 2 D_{0} + D_{-1})
 + N_{0}^{\pm},
\label{eqn-sdcc0} \\
\Delta_{-1}^{\pm} &= &
 \frac{1}{4(1+s)} (D_{-1})^{2}
 + \frac{(1+s)}{4} ( \Delta N_{-1} - \frac{1}{2} \Delta N_{0} )^{2}
 \nonumber \\ & & \mbox{ }\
 \pm \frac{1}{4} ( 2 \Delta N_{-1} - \Delta N_{0} ) D_{-1}
 + N_{-1}^{\pm}
\label{eqn-sdcc1}
\end{eqnarray}
Note that the second equation is independent of $\xi_0$. However, there
is an explicit dependence on the interaction strength $s$ and only for
$s=0$ do we recover the result of the Hubbard model.

This $s$ dependence can be understood by realizing that for the zero sector
and $-1<s<0$ the Bethe-Ansatz equations of the rapidities
${\bf k}_{-1}=(\lambda_{1}, \ldots, \lambda_{N_{-1}})$
are essentially the Bethe-Ansatz equations of the Heisenberg-Ising model.
The effect of the Bethe-Ansatz equations for the pseudo-momenta is simply
a renormalization.
Following Ref.\onlinecite{yy} we parametrize the anisotropy in the
Heisenberg-Ising model by $\Delta=- \cos(\mu)$. Then the correspondence is
established by setting $\mu =-\pi s$. Thus we may say that the behavior of
the spin wave exitations changes from ferromagnetic at $s\rightarrow -1^{+}$
($\Delta=1$) to antiferromagnetic at $s\rightarrow 0^{-}$ ($\Delta=-1$).
Furthermore, we expect to see free spin waves at $s=-1/2$.
This picture has been confirmed by a study of the transport properties of the
SC-model which we present in another publication.\cite{rstp}

An integral equation can also be given for $\xi_0$,
\begin{equation}
 \xi_0(k) = 1
  +\frac{1}{2\pi} \int_{-B}^{B} \xi_0(h) \theta'(h-k) {\rm d}h,
\end{equation}
where the kernel is as in equation (\ref{eqn-kern}).
Alternatively, we may simply express $\xi_{0}$ in terms of
thermodynamical response functions as
\begin{equation}
\xi_0^{2} = \pi v_{0} \kappa_{0} d_{0}^{2} = \pi d_{0}/v_{0}.
\label{eqn-xicc}
\end{equation}

\subsection{correlation functions and conformal expansion}

Given the conformal weights, we now construct the asymptotic
expressions for correlation functions.
For $-1<s<0$, we want to consider the following set of correlators:
Let $\psi_{\sigma}(x,t)$ denote the field operator of a particle
with spin $\sigma$. Later, we will additionally restrict the statistics
to be either bosonic or fermionic by restricting the possible values of
the pair ${\bf D}$.
Then the field correlator --- also called the one-particle reduced
density matrix --- is given by
\begin{equation}
C_{\psi}(x,t) =
 \langle \psi_{\downarrow}(x,t) \psi_{\downarrow}^{\dagger}(0,0)\rangle .
\label{eqn-psipsi}
\end{equation}
Defining the number operator
$n(x,t)= n_{\uparrow}(x,t) + n_{\downarrow}(x,t)$, we write the
density-density correlator
\begin{equation}
C_{n}(x,t) =
 \langle n(x,t) n(0,0) \rangle.
\label{eqn-nn}
\end{equation}
The spin-spin correlation functions are
\begin{equation}
C_{\sigma}^{z}(x,t) =
 \langle S^{z}(x,t) S^{z}(0,0) \rangle,
\label{eqn-long}
\end{equation}
\begin{equation}
C_{\sigma}^{\perp}(x,t) =
 \langle S^{-}(x,t) S^{+}(0,0) \rangle,
\label{eqn-trans}
\end{equation}
where we used $S^{z}= (n_{\uparrow}-n_{\downarrow})/2$ and $S^{+}=
\psi_{\uparrow}^{\dagger}\psi_{\downarrow}$.
Note that for systems that are rotationally invariant, such as the
Hubbard model in zero magnetic field, these two spin-spin correlators are
closely related, i.e.\ $C_{\sigma}^{z} = 2 C_{\sigma}^{\perp}$.

Following Ref. \onlinecite{fk90}, we also consider the correlation
function for singlet pairs,
\begin{equation}
C_{sing}(x,t) =
 \langle \psi_{\uparrow}^{\dagger}(x,t) \psi_{\downarrow}^{\dagger}(x,t)
         \psi_{\uparrow}(0,0) \psi_{\downarrow}(0,0)
 \rangle.
\label{eqn-sing}
\end{equation}
Note that all these correlators are of the form
$\langle A(x,t) A^{\dagger}(0,0)\rangle$.
By standard arguments of conformal field theory,\cite{cft}
we may deduce the leading terms and the critical exponents of the
long-distance behavior of these correlators by expanding $A$ in terms of
the primary fields $\phi_{\pm}$ while minimizing with respect
to ${\bf D}$ at the corresponding values of $\Delta{\bf N}$. Here the
above mentioned restrictions on ${\bf D}$ will become crucial.
This approach, however, will leave the expansion coefficients
undetermined and at special points in the phase diagram, they may
even vanish.

For $s>0$, the model exhibits a gap for breaking of pairs and
there are no spin waves.
Therefore the correlators (\ref{eqn-psipsi}), (\ref{eqn-long}) and
(\ref{eqn-trans}) will decay exponentially.
Let us introduce the pair field operator $\Psi$.
The pair density - pair density correlator can be written in terms
of the pair number operator $p= \Psi^{\dagger}\Psi$ as
\begin{equation}
C_{p}(x,t)= \langle p(x,t) p(0,0) \rangle
\label{eqn-p}
\end{equation}
and the pair field correlator is given by
\begin{equation}
C_{\Psi}(x,t)= \langle \Psi^{\dagger}(x,t) \Psi(0,0) \rangle.
\label{eqn-Psi}
\end{equation}
As before, we can construct these correlators by an expansion in
primary fields, minimizing with respect to $\Delta N_{1}$ and $D_{1}$.

%
%

\section{Asymptotics of the Correlation Functions for the Unbound Case}
\label{sec-acfa}

Due to the restrictions (\ref{eqn-srbcc}) and (\ref{eqn-srfcc}) on the
quantum numbers of a given state, the numbers
${\bf D}= (D_{0}, D_{-1})$ are integers or half-odd integers depending on the
parities of the pair $\Delta{\bf N}= (\Delta N_{0}, \Delta N_{-1})$ and
the statistics of $\psi^{\dagger},\psi$.
In particular, for fermionic particles we have
\begin{equation}
  D_{0} = \frac{\Delta N_{0}+\Delta N_{-1}}{2}\ ({\rm mod}\ 1), \mbox{  }\
  D_{-1}= \frac{\Delta N_{0}}{2}\ ({\rm mod}\ 1).
\label{eqn-srfdcc}
\end{equation}
We can now apply the scheme for calculating the leading asymptotic behavior
of the correlation function as outlined in the last section.
Following our selection rules, we therefore have for a fermionic system
\begin{equation}
\begin{array}{lll}
C_{\psi}:
 & \Delta N_{0}=1;                 & \Delta N_{-1}=1; \\
 & D_{0}= 0, \pm 1, \ldots;        & D_{-1}= \pm\frac{1}{2}, \ldots; \\
C_{n}:
 & \Delta N_{0}=0;                 & \Delta N_{-1}=0; \\
 & D_{0}= 0, \pm 1, \ldots;        & D_{-1}= 0, \pm 1, \ldots; \\
C_{\sigma}^{z}:
 & \Delta N_{0}=0;                 & \Delta N_{-1}=0; \\
 & D_{0}= 0, \pm 1, \ldots;        & D_{-1}= 0, \pm 1, \ldots; \\
C_{\sigma}^{\perp}:
 & \Delta N_{0}=0;                 & \Delta N_{-1}=1; \\
 & D_{0}= \pm\frac{1}{2}, \ldots;  & D_{-1}= 0, \pm 1, \ldots; \\
C_{sing}:
 & \Delta N_{0}=2;                 & \Delta N_{-1}=1; \\
 & D_{0}= \pm\frac{1}{2}, \ldots;  & D_{-1}= 0, \pm 1, \ldots\ .
\end{array}
\end{equation}
This is identical to the results for the repulsive Hubbard model, and
as in Ref. \onlinecite{fk90}, we will write the critical exponents
as functions of $\theta\equiv 2 \xi_0^{2}$. However, there is an additional
interaction strength dependence in the correlation functions due to
the explicit appearence of $s$ in equation (\ref{eqn-sdcc1}).
This is a novel feature and not true in the Hubbard model.
It emphasizes the close correspondence of the Heisenberg-Ising model and the
SC-model for $-1<s<0$ in the zero sector.

Following the scheme outlined briefly in the last section, we calculate the
leading asymptotics of the fermionic field correlator in the SC-model to be
\begin{eqnarray}
C_{\psi}(x,t)
 &\sim &
 \frac{1}{|x+iv_0 t|^{1/\theta+\theta/16}
          |x+iv_{-1} t|^{\frac{1}{2}+s^2/4(s+1)}}
  {\rm Re}\left[ A_0 e^{-ik_f x}
   \left(\frac{x+iv_{0} t}{x-iv_{0} t}\right)^{\frac{1}{4}}
   \left(\frac{x+iv_{-1} t}{x-iv_{-1} t}\right)^{\frac{1}{4}}
  \right] \nonumber \\ & &
 +
 \frac{1}{|x+iv_0 t|^{1/\theta+9\theta/16}
          |x+iv_{-1} t|^{\frac{1}{2}+s^2/4(s+1)}}
  {\rm Re}\left[ A_1 e^{-i3k_f x}
   \left(\frac{x+iv_{0} t}{x-iv_{0} t}\right)^{\frac{3}{4}}
   \left(\frac{x+iv_{-1} t}{x-iv_{-1} t}\right)^{\frac{1}{4}}
  \right]. \label{eqn-psif}
\end{eqnarray}
The density-density correlator is given by
\begin{eqnarray}
C_{n}(x,t)
 &\sim &
 n_0^2
 +
 A_1
 \frac{\cos(2k_f x + \Phi_1)}{|x+iv_{0} t|^{\theta/4} |x+iv_{-1} t|^{1/(1+s)}}
 +
 A_2
 \frac{\cos(4k_f x + \Phi_2)}{|x+iv_{0} t|^{\theta}}
 \nonumber \\ & &
 +
 A_3
 \frac{x^{2}-(v_{0}t)^{2}}{[x^{2}+ (v_{0}t)^{2}]^{2}}
 +
 A_4
 \frac{x^{2}-(v_{-1}t)^{2}}{[x^{2}+ (v_{-1}t)^{2}]^{2}}, \label{eqn-nnf}
\end{eqnarray}
and since the selection rules for the density-density correlator are identical
to the selection rules for the longitudinal spin-spin correlator, the above
calculation holds for $C_{\sigma}^{z}$ with different constants and the
replacement of ${\cal M}^2$ for $n_0^2$. Finally, for the transverse spin-spin
and the single-particle correlator we find
\begin{eqnarray}
C_{\sigma}^{\perp}(x,t)
 &\sim &
 A_0
 \frac{\cos(2k_f x + \Phi)}{|x+iv_{0} t|^{\theta/4} |x+iv_{-1} t|^{(1+s)}}
\nonumber \\ & &
 +
 \frac{1}{|x+iv_{-1}t|^{2+s^2/(1+s)}}
 {\rm Re}\left[A_1
  \frac{x+iv_{-1}t}{x-iv_{-1}t}
 \right], \label{eqn-lssf} \\
C_{sing}(x,t)
 &\sim &
 A_0
 \frac{1}{|x+iv_{0} t|^{4/\theta} |x+iv_{-1} t|^{1/(1+s)}} \nonumber \\ & &
 +
 \frac{1}{|x+iv_{0}t|^{4/\theta+\theta/4}}
 {\rm Re}\left[ A_1 e^{-i2k_f x}
  \frac{x+iv_{0}t}{x-iv_{0}t}
 \right]. \label{eqn-spf}
\end{eqnarray}
Following equation (\ref{eqn-xicc}), we calculate $\xi_0$ from the
Fermi velocity $v_{0}$.
In Fig.~\ref{fig-xicc}, we plot the lines of constant $\xi_0$ in the
$(d_{0}, s)$ plane.
Note that the value of $\theta(\xi_{0})$ at zero density is given by $2(1)$,
whereas for finite densities and vanishing
interaction strength $s\rightarrow 0^{-}$, we have
$\theta\rightarrow 4$ $(\xi_{0}\rightarrow \sqrt 2)$.
As expected, this is the same behavior as in the Hubbard model for
vanishing on-site interaction strength $u$.
In particular, the explicitely $s$-dependent exponents in the SC-model reduce
to constant values as $s\rightarrow 0^{-}$ which are equal to the corresponding
exponents in the Hubbard model.
However, we can not bound
$\theta$ between those two values as we could for the Hubbard model.
In fact, $\theta$ is larger than $4$ and continues to increase for finite
densities and increasing negative interaction strength $s\rightarrow -1^{+}$.
A plot of $\theta$ as a function of the density $d_{0}$
for different values of the interaction strength $s$ is given in
Fig.~\ref{fig-xidcc}.

For bose statistics, $D_{0}$ and $D_{-1}$ are restricted to integer values.
The correlators of diagonal operators, i.e.\ the density-density
correlator $C_{n}$ and the longitudinal spin-spin correlator $C_{\sigma}^{z}$
are independent of statistics, and so only the correlators $C_{\psi}$,
$C_{\sigma}^{\perp}$ and $C_{sing}$ change.
We find for their asymptotics
\widetext
\begin{eqnarray}
C_{\phi}(x,t)
 &\sim &
 A_{0} \frac{1}{|x+i v_{0}t|^{1/\theta} |x+i v_{-1}t|^{(1+s)/4}} \nonumber \\ &
&
+ \frac{1}{|x+i v_{0}t|^{\theta/4+1/\theta}|x+i v_{-1}t|^{(s^2+4s+10)/8(1+s)}}
\times \\ & &
  \mbox{ }\ {\rm Re}\left[ A_{1} e^{- i 2 k_{f} x}
    \left( \frac{x+iv_{0}t}{x-iv_{0}t} \right)^{1/2}
    \left( \frac{x+iv_{-1}t}{x-iv_{-1}t} \right)^{1/2}
  \right], \label{eqn-psib}\\
C_{\sigma}^{\perp}(x,t)
 &\sim &
 A_{0} \frac{1}{|x+i v_{-1}t|^{(1+s)}} \nonumber \\ & &
+ \frac{1}{|x+i v_{0}t|^{\theta/4}|x+i v_{-1}t|^{2+s^2/(1+s)}}
  {\rm Re}\left[ A_{1} e^{- i 2 k_{f} x}
    \left( \frac{x+iv_{-1}t}{x-iv_{-1}t} \right)
  \right], \label{eqn-tssb}\\
C_{sing}(x,t)
 &\sim &
 A_{0} \frac{1}{|x+v_{0}t|^{4/\theta}} \nonumber \\ & &
+ \frac{1}{|x+i v_{0}t|^{\theta/4+4/\theta}|x+i v_{-1}t|^{1/(1+s)}}
  {\rm Re}\left[ A_{1} e^{- i 2 k_{f} x}
    \left( \frac{x+iv_{0}t}{x-iv_{0}t} \right)
  \right]. \label{eqn-singb}
\end{eqnarray}
\narrowtext

%
%

\section{Asymptotics of the Correlation Functions for the Bound Case}
\label{sec-acfr}

Due to the restriction (\ref{eqn-src}) on the quantum numbers of a given
state, $D_{1}$ is an integer or half-odd integer depending on the
parity of $\Delta N_{1}$ for both bose and fermi statistics of the
particles, i.e.\
\begin{equation}
D_{1} = \frac{\Delta N_{1}}{2}\ ({\rm mod}\ 1)
\end{equation}
This selection rule is just the same as the case of one-component
bosons, and so we find for the asymptotics of the pair density correlator
\begin{eqnarray}
C_{p}(x,t) - d_{1}^{2}
 &\sim &
 A_{1} \frac{x^{2}-(v_{1}t)^{2}}{[x^{2}+ (v_{1}t)^{2}]^{2}} \nonumber \\ & &
+ A_{2} \cos(2 k_{f} x + \varphi_{1}) \frac{1}{|x + i v_{1}t|^{\theta}}
\end{eqnarray}
and for the pair field correlator
\begin{eqnarray}
C_{\Psi}(x,t)
 &\sim &
 A_{1} \frac{1}{|x + i v_{1}t|^{1/\theta}} \nonumber \\ & &
+ \frac{1}{|x+i v_{1}t|^{\theta+1/\theta}}
  {\rm Re}\left[ A_{2} e^{i 2 k_{f} x} \frac{x-iv_{1}t}{x+iv_{1}t}
  \right].
\label{eqn-Psi-exp}
\end{eqnarray}
Here we again defined an exponent $\theta= 2 \xi_{1}^{2}$.
Following equation (\ref{eqn-xic}), we can calculate $\xi_1$ from the
Fermi velocity of pairs $v_{1}$.
In Fig.~\ref{fig-xic}, we plot the lines of constant $\xi_1$ in the
$(d_{0}, s)$ plane.
Note that the value of $\theta(\xi_{1})$ at zero density is given by $8(2)$,
whereas for finite densities and vanishing interaction strength
$s\rightarrow 0^{+}$, we have $\theta\rightarrow 4$ $(\xi_{1}\rightarrow
\sqrt 2)$.
A plot of $\xi_1$ as a function of the density $d_{0}$
for different values of the interaction strength $s$ is given in
Fig.~\ref{fig-xidc}.

%
%

\section{The Non-Interacting Two-Component System}
\label{sec-nitcs}

At $s=0$, the system reduces to a non-interacting two-component gas
and we may expect a certain continuity in the behavior of the
correlators at this point.
Indeed, as $s\rightarrow 0^{-}$, the two Fermi velocities $v_{0}$ and
$v_{-1}$ both approach the Fermi velocity of a non-interacting
single-component model, i.e.\ $v_{0}(s\rightarrow 0^{-})=
v_{-1}(s\rightarrow 0^{-})= \pi d_{0}/2$.
Consequently, the correlation functions of the bosonic (fermionic)
system reduce to the correlation functions of a non-interacting bose
(fermi) system with two components, i.e.\ with half the one-component
fermi momentum.
Using the language of conformal field theory, we can thus describe
the excitations of the non-interacting two-component gas by a
$c=2$ generalized Gaussian model.\cite{w89}

 From the expression of the dressed charges $\xi_{0}$ and $\xi_{1}$,
we see that $\xi_{1}^{2}= \frac{1}{2} \xi_{0}^{2} \frac{v_{0}}{v_{1}}$.
As $s\rightarrow 0^{+}$, the Fermi velocity of the pairs goes to the
Fermi velocity of a one-component free bose gas with doubled particle
mass, i.e.\
$v_{1}(s\rightarrow 0^{+})=\pi d_{1}/2=\frac{1}{2} v_{0}(s\rightarrow 0^{-})$.
Therefore, we expect $\theta_{1} = \theta_{0}$ at $s=0$ and this is indeed
true as shown above.
Furthermore, the free energy of the system should be uniquely specified
at $s=0$. Following (\ref{eqn-fe}) we may write the finite temperature
corrections for the unbound case as
\begin{equation}
F(T) \simeq F(T=0) - \frac{\pi T^2}{6}
 \left( \frac{1}{v_{-1}} + \frac{1}{v_{0}} \right),
\label{eqn-fecc}
\end{equation}
whereas for the bound case we have
\begin{equation}
F(T) \simeq F(T=0) - \frac{\pi T^2}{6 v_{1}}.
\label{eqn-fec}
\end{equation}
As predicted, these two equations are in agreement at $s=0$ and identical
to the free energy of a non-interacting $c=2$ system.

The bound pairs for $s>0$ are singlets. Therefore we might expect that
the pair field correlator (\ref{eqn-Psi}) becomes identical to the
singlet pair correlators (\ref{eqn-sing}) and (\ref{eqn-singb})
of the unbound case as $s\rightarrow 0$.
However, $\Psi^{\dagger}$ creates pairs with characteristic length
scale $1/s$ and not just two particle wave functions.
Thus, the pair wave functions include a normalization
factor $\sqrt s$. As $s\rightarrow0^{+}$, the leading terms of the
conformal expansion (\ref{eqn-Psi-exp}) consequently vanish and higher
order terms become important. It should therefore come as no surprise
that the expansions (\ref{eqn-sing}), (\ref{eqn-singb}) and
(\ref{eqn-Psi-exp}) do not agree at $s=0$.



\begin{figure}
  \caption{Lines of constant universal behavior for the unbound case.
  Contours of constant value of the dressed charge $\xi_{0}$ in the
  $(d_{0},s)$ plane are shown. The lines represent increments of $.2$
  starting from $\xi_{0}=1.0$ at $d_{0}=0$ up to $\xi_{0}=1.8$.
  The dashed line correspond to the value $\xi_{0}= \protect\sqrt 2$ of a
  non-interacting system.
  \label{fig-xicc}}
\end{figure}
\begin{figure}
  \caption{Plot of $\theta$ as function of particle density $d_{0}$ for
  various values of interaction strength $s$ for the unbound case.
  \label{fig-xidcc}}
\end{figure}

\begin{figure}
  \caption{Lines of constant universal behavior for the bound case.
  Contours of constant value of the dressed charge $\xi_{1}$ in the
  $(d_{0},s)$ plane are shown. The lines represent increments of $.2$
  starting from $\xi_{1}=2.0$ at $d_{0}=0$ down to $\xi_{1}=1.2$.
  \label{fig-xic}}
\end{figure}
\begin{figure}
  \caption{Plot of $\theta$ as function of particle density $d_{0}$ for
  various values of interaction strength $s$ for the bound case.
  \label{fig-xidc}}
\end{figure}

\end{document}